# BeeCup: A Bio-Inspired Energy-Efficient Clustering Protocol for Mobile Learning

Feng Xia[1], Xuhai Zhao[1], Jianhui Zhang[2], Jianhua Ma[3], Xiangjie Kong[1*]

[1]School of Software, Dalian University of Technology, Dalian 116620, China;
Email: f.xia@ieee.org

[2]Institute of Computer Application Technology, Hangzhou Dianzi University, Hangzhou 310018, China;
Email: jhzhang@ieee.org

[3]Faculty of Computer & Information Sciences, Hosei University, Tokyo 184-8584, Japan;
Email: jianhua@hosei.ac.jp

**Abstract:** Mobile devices have become a popular tool for ubiquitous learning in recent years. Multiple mobile users can be connected via ad hoc networks for the purpose of learning. In this context, due to limited battery capacity, energy efficiency of mobile devices becomes a very important factor that remarkably affects the user experience of mobile learning. Based on the artificial bee colony (ABC) algorithm, we propose a new clustering protocol, namely BeeCup, to save the energy of mobile devices while guaranteeing the quality of learning. The BeeCup protocol takes advantage of biologically-inspired computation, with focus on improving the energy efficiency of mobile devices. It first estimates the number of cluster heads (CHs) adaptively according to the network scale, and then selects the CHs by employing the ABC algorithm. In case some CHs consume energy excessively, clusters will be dynamically updated to keep energy consumption balanced within the whole network. Simulation results demonstrate the effectiveness and superiority of the proposed protocol.

## 1. Introduction

The ubiquitous deployment of wireless technologies has revolutionized the way of knowledge acquisition. Nowadays, learning is no longer limited in classroom with a teacher ahead. Mobile learning (*m-learning*), which facilitates education through wireless networks, provides learners with an opportunity to access abundant teaching materials and educational services using hand-held mobile devices, such as smart phones, laptops and so on. According to Gartner, a market research institution, the total number of mobile phones had reached to 1.211 billion in 2010 in the world [1] and the sales volume of smart phone had surpassed desktop machines to become the most prevalent computing platforms in 2011 [2]. As the mobile devices are available to more and more people, the trend of the development of m-learning will be inevitable.

M-learning emphasizes its characteristics on "moving in" learning, contextual relevance, and other new features [3][4]. This benefits learners to learn at any time, any place as long as the mobile devices can access the server from internet. However, as device display, CPU and wireless network interface card drain much energy when displaying and processing the intensive multimedia content [5], which may shorten the study time and significantly decrease the quality of learning. Accordingly, how to save the energy and provide the learners with long and persistent acceptable quality of learning experience becomes a great challenge.

It has been realized that the approach which is likely to effectively provide energy-efficient [6][7] and load balancing [8] solution is by using the hierarchical structure, which is also called clustering. Clustering can be extremely effective in one-to-many, many-to-one, one-to-any, or one-to-all (broadcast) communication. For example, in one-to-many communication, the teacher only needs to transmit the multimedia streaming or other teaching files to the Cluster Heads (CHs) that have more residual energy and less mobility. Then the CHs dispatch the teaching resources to their Regular Nodes (RNs) within the communication range of the clusters, and the RNs only need to communicate with their CHs and consequently can omit the long distance communication with the server. In this way, the clustering structure can effectively save the mobile devices' energy.

---

[*] Corresponding author: Xiangjie Kong; Email: xjkong@ieee.org

However, the existing clustering protocols can not be directly applied to the mobile learning because some protocols (e.g. [9], [10]) cannot determine the number of clusters adaptively according to the varying network environments. Some of them don't consider the maximal cluster size which may overpass the upper limit of the communication methods [11]. Some clustering protocols may create many single-node clusters which have to communicate with the server node directly, thus consuming much energy.

In this paper we design a new energy-efficient clustering protocol (called BeeCup) for m-learning applications over ad hoc networks. The BeeCup protocol takes advantage of bio-inspired computation. It firstly estimates how many clusters are suitable for the current m-learning network, and then selects some mobile devices as CHs taking into account several factors affecting the energy consumption as well as learning quality in practice. By introducing the clustering into m-learning scene, only the devices which are selected as CHs will communicate with the server directly using long-distance communication protocol, e.g. WLAN. Other users who want to download resources or submit the feedback only need to communicate with their CHs which are near and hence low-power communication methods can be employed. This can substantially save the mobile devices' energy, thus prolonging the learning time and enhancing the learning experience.

In order to select the most suitable mobile devices to serve as CHs while considering factors that remarkably affect the energy consumption, we need to model the clustering in m-learning. It has been shown in many papers that the collective behavior of social insects has many attractive features [12]. Inspired by the food foraging behavior of bees [13], Karaboga *et al.* [9] proposed the bio-inspired artificial bee colony (ABC) algorithm for solving complex optimization problems. ABC is a population-based meta-heuristic approach which has very good convergence rate and robustness, and has been applied in many fields such as numerical function optimization [14], clustering [9] and so on. Here we will use the ABC algorithm to estimate the number of clusters for the current mobile learning scene, and to optimize the fitness function which covers the factors that have close connection with the energy consumption of mobile devices such as their residual energy, real-time mobility and so on. In our previous work [15], we proposed the EBABC protocol, a clustering protocol for the ad-hoc sensor networks. In this paper, we substantially extend our previous work [15]. For instance, we will improve the process of CH number estimation by using a new structure of solution for ABC, thus making this process faster. We also propose two flexible clustering maintenance methods.

The major contributions of this paper can be summarized as follows:

1) We introduce the clustering protocol into mobile learning to prolong the network lifetime by saving mobile devices' energy, and to enhance the quality of experience as well as learning effects. The proposed BeeCup protocol explores bio-inspired computation to adaptively determine the most proper number of clusters instead of assigning one manually.

2) We find a suitable fitness function for the scene of mobile learning considering factors affecting the energy consumption and learning quality. We also propose two intelligent clustering maintenance methods to adjust the clusters dynamically to achieve load balance among the networked mobile devices during system runtime.

3) We conduct extensive simulations to evaluate the performance of the proposed BeeCup protocol under different scenarios, as compared against three existing well-known protocols. Results are presented to show the effectiveness and superiority of the BeeCup protocol.

The rest of this paper is organized as follows. Section 2 briefly surveys related work. Section 3 outlines the problem statement and enumerates our assumptions. Section 4 presents the BeeCup protocol and the maintenance mechanisms. Section 5 shows their effectiveness via simulations, and compares it to other clustering techniques. Finally, Section 6 gives concluding remarks.

## 2. Related Work

Many previous studies have investigated techniques to reduce energy consumption in m-learning, with focus on the following aspects: saving energy by personalizing the multimedia content [5][16], saving energy at the playing stage [17][18], using optimized transport method during the transmission of teaching resource [19][20][21] and so on.

The BAAMLS system [5] varied the parameters of a multimedia clip such as the encoding technique and video resolution based on the mobile device's battery and its behavior when the residual energy was not sufficient for playing the requested media content. This decreases power consumed by the mobile device to retrieve, decode and display the multimedia content while maintaining a good perceived quality and prolongs the time of learning. In BaSe-AMy [16], the remaining battery level and video stream duration as well as the packet loss rate were assessed to dynamically adjust the bit-rate of the stream and display brightness which could save energy, bring a good quality of learning experience and prolong the network lifetime. However, the above systems may affect the learners' quality of experience with less resolution or low bit-rate of stream.

Displays have been known as one of the major power consumers in mobile systems and different display technologies have different energy consumption profiles [17]. Effective control of the display is very important for energy saving during mobile learning. Cheng *et al*. [22] proposed an algorithm for Liquid Crystal Display (LCD) screen devices to adapt the backlight of the screen while also adjusting the intensity level of each pixel in the video frame to compensate for the degradation of video quality. The GreenVis scheme proposed in [18] builds a multi-objective optimization approach based on the Organic Light-Emitting Diode (OLED) power model to find energy-saving sequential color schemes and reduce power consumption on OLED display.

Some works focus on the optimization of the communication protocols, such as CoolSpots [21] and SwitchR [20] used the Bluetooth as communication method when the data rate did not reach the Bluetooth limit. Bluetooth was also used to wake up the WLAN channel when necessary. Nevertheless, these methods did not consider the whole network condition and could not prolong the network lifetime by cooperation.

None of the above works take the network structure into consideration. It is realized that hierarchical structure is likely to provide a scalable and energy-efficient solution. Many clustering algorithms have been proposed for the purpose of energy efficiency. In WSNCABC [9], distance and the residual energy of the cluster heads were considered for clustering using the ABC algorithm. However, it did not consider the coverage of the CHs which may leave many single-node clusters. The negative impacts of single-node cluster on the network performance have been shown in [24]. The CONET protocol [11] reformed clusters according to each node's bandwidth requirement, energy use, and application type, using Bluetooth as intra-communication method and WiFi for inter-cluster communication. The maximum number of regular nodes in a cluster is not specified which may overpass the limit of the communication protocol [26]. In LEACH clustering protocol [10], the CHs were selected randomly and every node could be chosen as CH in a round with a local determined probability. A node became a CH for the current round if a random between 0 and 1 was less than the threshold showed in Eq. (1):

$$T(s) = \begin{cases} \dfrac{p}{1 - p \times (r \bmod \dfrac{1}{p})} & if \quad s \in G \\ 0 & otherwise \end{cases} \quad (1)$$

where $p$ is the predefined percentage of CHs, $r$ is the number of current round and $G$ is the set of nodes that have not been chosen as CH in the last $1/p$ rounds. A prerequisite for LEACH is that the nodes must be homogenous, that may limit the scope of application. The SEP [28] improved the LEACH and could be applied to the heterogeneous network in which some of the nodes may equip with extra energy. Supposing *m* is the fraction of advanced nodes and $\alpha$ is the addition energy factor between advanced nodes and normal nodes, the threshold for normal nodes and advanced nodes to be CHs took the two parameters into consideration. By assigning different thresholds for normal and advanced nodes, the two kinds of nodes have different probability to become CHs.

There are also some clustering methods for energy efficient based on bio-inspired or heuristic algorithm in which the clustering were considered as resource-constrained optimization problems, such as the Tabu search algorithm [23] and Particle Swarm Optimization (PSO) [29][30]. The authors of [23] used the Tabu search to seek an energy-optimal topology that maximizes network lifetime while ensuring simultaneously full area coverage and sensor connectivity to cluster heads. The PSO-GA [10] utilized GA to choose the cluster heads and their number, and then used PSO to select the clusters' members based on the combined factors of distance, energy and the CH's load. However, it didn't consider the maximum size of a cluster and the CHs' coverage which may leave many single-nodes. It had been proved in [31] that the ABC optimization algorithm has better performance for solving constrained optimization problems.

Unlike previous work about energy efficient in mobile learning, our BeeCup protocol introduces the clustering into mobile learning to save the mobile devices' energy and to enhance the quality of experience and learning effects. We use the ABC algorithm to adaptively determine the number of clusters rather than manually assign one. We also propose a suitable fitness function for the scene of mobile learning considering the factors of distance, mobility, energy and number of single-node cluster. Besides, our proposed protocol also includes two intelligent cluster maintenance methods to adjust the clusters dynamically before re-clustering.

**3 Problem Statement**

Wireless networking is the main component of m-learning. In this paper, we use the WLAN and Bluetooth as basic ways of communication for the both-way communication between the server and learners. The cluster heads use WLAN to communicate with the server while using Bluetooth as intra-cluster communication method. The schematic diagram of clustered mobile learning scene can be seen in Fig. 1.

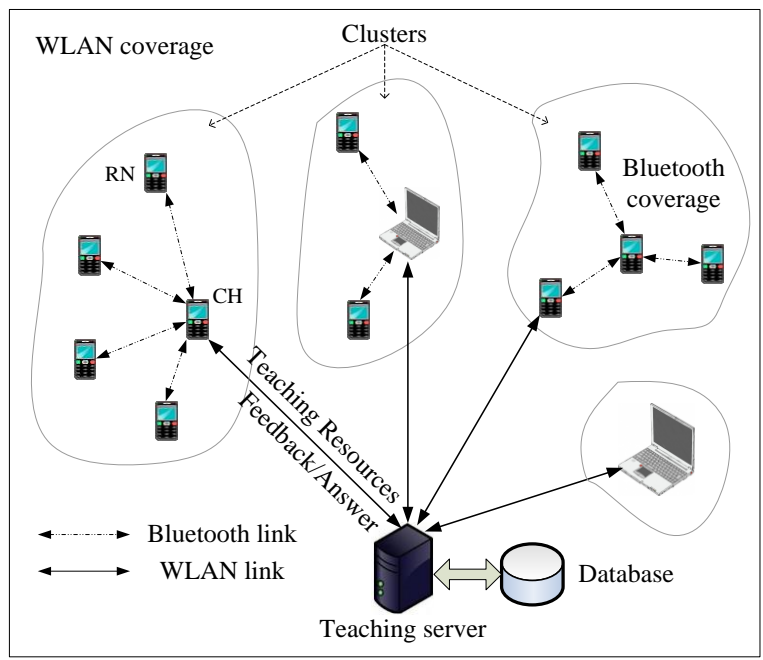

Figure 1. Clustered m-learning schematic diagram

Fig. 1 is a diagrammatic sketch of electronic classroom. Students can use the mobile devices such as smartphones or laptops to learn the materials they are interested in from the resource server and could also interact with the teacher. The teaching server uses the WLAN which has a long communication radius to communicate with the CHs elected by our clustering protocol. Within the cluster, the RNs use the Bluetooth to communicate with their CHs.

During a learning session, cameras and microphones may be used to record the live scenes of the "classroom" and transmit the videos, audio along with some other files to the learners who are served as CHs using the wireless technology of WLAN. Then the CHs broadcast the learning resources to its members in the cluster. In the meantime, the instructor may receive questions, feedback and reports from the learners through the converse way.

Here we describe the m-learning scene as follows:
1) The mobile device is treated as a node which can move at a low speed and with limited battery.
2) Each node can use both WLAN and Bluetooth to communicate.
3) The server is seen as a base station which contains much teaching resource.
4) The server has infinite energy and has the ability of powerful computing so that we can run our protocol on the server.

Our main design goal is to divide the network into an optimal number of clusters considering the particular factors of mobile learning such as mobility, residual energy of the mobile devices, and distance to the server and so on. So some definitions and constraints of cluster are necessary for the process of clustering.

The network is modeled on a hyper graph with a node set $V= \{1, 2… n\}$. The clusters $C= \{C_1, C_2... C_k\}$. Additionally, let us define

- $CH_i$: the cluster head of cluster $C_i$
- $S_i$: the size of cluster $C_i$
- $RN_{ij}$: the $j_{th}$ regular node of cluster $C_i$
- $A_{ij} = 1$: if cluster $C_i$ contains node $j$, 0 otherwise
- $D_{bc}$: distance between node $b$ and node $c$

Based on those definitions, all the clusters have to meet the following constraints:

$$(\bigcup_i C_i = V) \wedge (\bigcap_i C_i = \varnothing) \tag{2}$$

$$\sum_{i=1}^{k} A_{ij} = 1, \quad for \quad \forall j \in \{1,2,...,n\} \tag{3}$$

$$\sum_{j=1}^{S_i} A_{ij} \leq S_{max}, \quad for \quad \forall i \in \{1,2,...,k\} \tag{4}$$

$$D_{RN_{ij},CH_i} \leq R_B, \quad for \quad \forall i \in \{1,2,...,k\}, j \in \{1,..,S_i\} \tag{5}$$

$$N_s = \sum_{i=1}^{k} S_i, \quad for \quad S_i = 0 \tag{6}$$

Eq. (2) ensures that the clusters must be disjoined sets and none of two clusters have overlapped nodes. Eq. (3) represents each node must belong to only one cluster, and it is called single-node cluster if there is only one node which severs as CH in a cluster. In Eq. (4) $S_{max}$ stands for the maximum size of a cluster. And for the clusters using Bluetooth, the number of regular nodes could not exceed seven. Eq. (5) shows the radius of a cluster. A node can only join a cluster if the distance between the node and the CH is less than $r_B$ which is 10$m$ in our protocol. Eq. (6) defines the number of single-node cluster which has no regular nodes. As the CHs have to communicate with the base station directly using WLAN, we must decrease this number to reduce energy consumption of the nodes and prolong the network lifetime, which is also a significant factor in determining the number of CH.

We model the clustering in the mobile learning scene as an optimization problem. That means we will organize the mobile devices as clusters which fulfill the constraints above and take the factors of mobile devices' mobility($M$), residual energy($E$), distance of communication($D$) as well as number of single node clusters($S$) into consideration to find the near optimum clustering scheme.

$$CH\_Number\_Fitness = \omega_1 \times M + \omega_2 \times S \tag{7}$$

$$CH\_Select\_Fitness = \alpha \times M + \beta \times E + \gamma \times D + \delta \times S \tag{8}$$

Eq. (7) means that we have to take the mobile devices' velocity as well as the number of single nodes into consideration to roughly divide the whole network into several relatively small stable regions. Then, the second part is to select the qualified CHs for each small region according to the specific condition of the mobile devices as Eq. (8) defines.

Before describing our clustering protocol in detail, we make the following assumptions:

- Each node has a unique ID, known to the server and to the node itself.
- Each node has two communication methods: WLAN and Bluetooth. The radius of WLAN is signed as $R_W$, and the radius of Bluetooth is $R_B$.
- The size of the Bluetooth network is smaller than eight as the Bluetooth ad-hoc network is composed of one master device and up to seven slave devices [32].
- We use routing protocol for communication between node and base station, in which each node can communicate directly with the base station using WLAN.
- The node can get its location through using GPS. Nodes' location information is available to the base station before running the clustering protocol.
- The speed information is available at all times by using accelerometer [33][34]. The nodes can move at a low speed during the m-learning.
- The nodes have ideal sensing capabilities. That is to say the nodes can detect any event inside the sensing range but the quality of communication will fade with the increasing of the distance. Any event occurring outside R is not detected.

## 4 Protocol Design
### 4.1 Overview

In the previous section, we have modeled the clustering in mobile learning scene as an optimization problem. Some aspects have to be considered carefully to divide the network into proper clusters. Firstly, the number of clusters and communication methods should be self-adaptive according to the mobile devices' condition. Secondly, after determining how many clusters are proper for the network, the CHs have to be selected seriously considering the factors of energy, mobility and so on, to make the clusters stable and prolong the network lifetime.

Since CHs are very important, the number and selection of CHs have to be carefully determined. In BeeCup, we use the ABC algorithm to select CHs. In ABC algorithm the employed bees, onlooker bees and the scouts work together to look for the most qualified food source with short distance, rich food, easy to acquire and some other excellent characteristics. Each food source corresponds to a solution of an optimization problem. The scouts search for new food sources to expend the diversity of the solution. We will use this bio-inspired algorithm to find out the most suitable number of clusters and select the proper CHs for each cluster. The comparisons of clustering and bee colony foraging are as Table 1 shows.

Table 1. Analogy between clustering problem and ABC

| ABC | clustering problem |
|---|---|
| solution | a set of CHs |
| dimension of solution | number of CHs |
| fitness of solution | quality of selected CHs |
| global best solution | CHs of highest quality |

Each solution represents a clustering scheme of the mobile learning network and through constant optimization of ABC to find the near optimum solution. As for the fitness of solution, we will define two fitness functions for the cluster number determination and CHs selecting processes of our clustering protocol and take the mobile devices' condition into condition to select the CHs with highest quality.

Our BeeCup protocol contains three main parts. Firstly, estimate the number of CHs by considering the nodes'

mobility, the communication radiuses of WLAN and Bluetooth as well as the maximum cluster size. By dividing the network into *K* stable areas which have low relative velocity and proper number of nodes, we could realize the distribution of the network and how many clusters is suitable. The second part is to choose *K* CHs based on the following constraint: 1) the distances to CHs from their member nodes and to the base station from CHs; 2) the mobility so as to select *K* stable CHs; 3) the residual energy at the meantime and the number of single-node cluster which has to communicate with the base station directly through WLAN.

The BeeCup protocol also includes two cluster maintenance mechanisms during re-clustering intervals: one is locally re-clustering if a node is no longer appropriate for being a CH due to its low residual energy and the second one is to balance local load by adjusting the RNs. The flowchart of our Bee-cup clustering protocol is shown in Fig. 2.

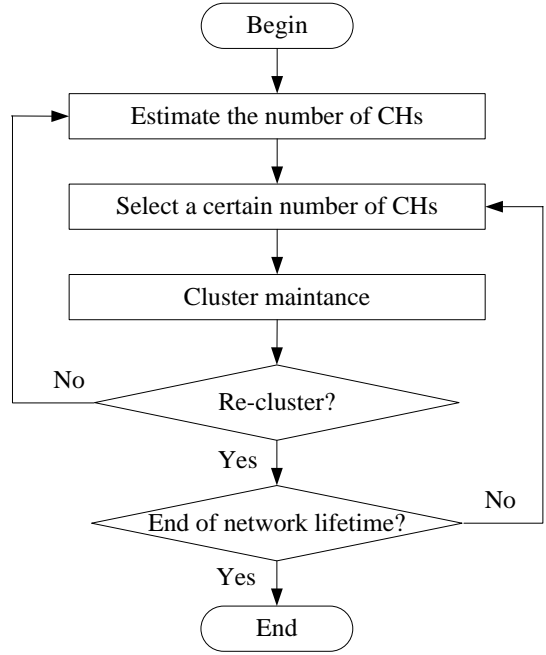

Figure 2. Flowchart of BeeCup protocol

**4.2 Number of Clusters**

Estimating the number of CHs is necessary before clustering the network especially for the randomly distributed network or network with irregular area. Some works (e.g. [11], [19]) manually assign a percentage of nodes to be CHs without fully considering the network condition and communication parameters such as radius and max cluster size. For example, different transmission powers or communication methods may have different communication radius which have essential influence on the coverage area.

Here, we define the fitness function as Eq. (9) using the ABC algorithm to find a solution with the highest fitness. $k$ represents the CHs from the non-empty cluster which will use the WLAN to communicate with the base station. $V_i$ stands for the instant velocity of node $i$. In order to let the value of the function as small as possible, we must limit the value of $k$ and choose the nodes with similar speed as CHs. $\omega_1$ and $\omega_2$ whose sum is 1 are used to control the weight of the two factors.

$$CHNumber\_Fitness = \omega_1 \times \frac{k}{n} + \omega_2 \times \frac{1.0}{\sum_{i=1}^{k} V_i + 1.0} \qquad (9)$$

As determining the number of CHs is the first part of our protocol which will have an important influence on the cluster formatting. Here we use ABC algorithm to optimize the number by considering the constraints Eqs. (2) - (6) as well as the nodes' mobility. Our goal is to estimate the optimum number of CH by dividing the network into several stable areas whose radius are less than $r_B$ and have no more than $S_{max}$ nodes.

As we know the goal of ABC algorithm is to find a solution with the highest fitness, which is implemented through *MCN* cycles. Here we set the parameter *D* as *n* which is the total number of nodes. A solution held by an employed bee has the following structure in Table 2. The index stands for the number of node and its value is 1 if it's a CH.

Table 2. The structure of solution

| Index | 1 | 2 | 3 | 4 | 5 | 6 | 7 | 8 | 9 | 10 | 11 | 12 | 13 | 14 | 15 | … | N |
|---|---|---|---|---|---|---|---|---|---|---|---|---|---|---|---|---|---|
| Value | 1 | 0 | 0 | 0 | 1 | 0 | 0 | 1 | 0 | 0 | 0 | 0 | 1 | 0 | 0 | … | 1 |

The algorithm of ABC is shown in Fig. 3 in which the fitness function Eq. (7) will be used.

```
Initialization
    Initialize clustering and ABC parameters
    FOR i=1 TO SN
        Initialize_CHNumber_Solution(i)
    End FOR
Repeat
    FOR i=1 TO SN
        Calculate_CHNumber_Fitness(i)
    End FOR
    [Send Employed Bees]
    FOR i=1 TO SN
        Produce neighbor solution using: v_{ij}=x_{ij}+φ_{ij}(x_{ij}-x_{kj})
        Calculate_CHNumber_Fitness(i)
        Greedy selection
    End FOR
    Calculate probabilities for each solution
    [Send Onlooker Bees]
    FOR i=1 TO SN
        Choose a solution depending on the probabilities
        Produce neighbor solution
        Calculate_CHNumber_Fitness(i)
        Greedy selection
    End FOR
    Memorize the best solution
    [Send Scout Bees]
        Replace the abandoned solution with new produced one
UNTIL termination condition is met
```

Figure 3. Estimating CH's number using ABC

The proposed algorithm firstly initializes the *SN* solutions by filling the above solution structure with 0 or 1 randomly as initial solutions, and then calculates the fitness of each employed bee's solution using the procedure shown in Fig. 4.

```
FOR j = 1 TO N
    FOR k = 1 TO CH_Number(i)
        IF Distance(j, CH[k]) < r_B AND RN_Number_of_CH(CH[k]) < S_max
            Add_RN_to_CH(j, CH[k])
        END IF
    END FOR
END FOR
S[i] = Calculate_Single_Node_Cluster(i)
p_1 = S[i], p_2 = 0
FOR k=1 TO CH_Number(i)
    p_2 += V[N[k]]
END FOR
RETURN ω_1×p_1/N + ω_2×1.0/(p_2+1.0)
```

Figure 4. Calculating fitness for CH number

The procedure of calculating the fitness for CH number includes two parts: Go through all the nodes of each solution structure and let the regular nodes in the solution join into the nearest CH within its communication range $r_B$, and the number of nodes must not overpass the maximum cluster size $S_{max}$; Then calculate the fitness of this solution concerning the single nodes of each cluster $S[i]$ and the CHs' velocity $V[N[k]]$ where $N[k]$ stands for the node number of the $k_{th}$ CH, and the CH_Number($i$) means the number of CHs of solution $i$.

At the start of each cycle in ABC, the employed bees go to their last memorized food source and try to find one nearby according to the algorithm, where $j$ is the index for a dimension, $i$ is the index of the current solution exploited by the employed bee and $k$ stands for a randomly chosen solution and has to be different from $i$. $\varphi_{ij}$ is a random number in [-1, 1] which controls the distance between the $j$'s factor between the solution $k$ and $i$. Through this way, the $j_{th}$ dimension of the solution is modified. The employed bee executes a greedy selection between the original solution and the new one and keeps the more excellent one in its memory.

The onlooker bees wait at the dancing zone of the nest and decide to select a food source from the information of employed bees according to the probability calculated by the Eq. (10):

$$Pi = \frac{fit_i}{\sum_{k=1}^{SN} fit_k} \tag{10}$$

where $fit_i$ and $fit_{best}$ stand for the solutions of $i_{th}$ employed and the best global solution respectively. When the onlooker bee gets the selected food source, it also tries to find a new one as the employed bees and selects the food source with higher fitness. After all the onlooker bees finish the above procedure, the global best solution is updated with the achieved best solution.

If a solution is not modified through *limit* cycles which are a predefined of ABC, the solution is abandoned and replaced with a new one found by the scout.

When the cycle reaches the maximum cycle number *MCN*, the number of clusters as well as the single-node clusters can be computed from the global best solution.

**4.3 Network Clustering**

After estimating how many clusters is well suitable for the current network condition, in this sub-section, we will select the *K* CHs using ABC algorithm based on the factors that mostly affecting the energy consumption of mobile devices such as distance, residual energy, mobility and the number of single-node clusters.

The solution structure of network clustering procedure is a vector whose dimension is *K* that represents the CHs to be found, through *MCN* cycles to constantly optimize the solutions of the employed bees.

Our fitness function is Eq. (11) which contains four parts illustrated above. We will use the result of the function to execute greedy selection process in ABC algorithm between two solutions. $α$, $β$, $γ$ and $δ$ are the weighing factors for the corresponding parameters. In case some parameters' orders of magnitude are not the same, we use the normalized parameters. We will give a detailed description to each parameter.

$$CHSelect\_Fitness = \alpha \times nD + \beta \times nE + \gamma \times nM + \delta \times nSN \tag{11}$$

The procedure of calculating the fitness for CH selecting is shown in Fig. 5. Similar to the first part of our protocol, constraints of Eqs. (2)-(6) have to be met. The regular nodes try to find the CHs between which the distance $Distance(j, CH[k])$ is less than $r_B$, and the size of clusters could not overpass the maximum cluster size $S_{max}$. The $p_1$, $p_2$, $p_3$ and $p_4$ stand for the four parameters to be considered during the clustering and their calculating functions will be discussed one by one.

```
FOR j = 1 TO N
    FOR k = 1 TO CH_Number(i)
        IF Distance(j, CH[k]) < r_B AND RN_Number_of_CH(CH[k]) < S_max
            Add_RN_to_CH(j, CH[k])
        END IF
    END FOR
END FOR
p_1 = Calculate_Single_Node_Cluster(i)
p_2 = Calculate_Mobility(i)
p_3 = Calculate_Distance(i)
p_4 = Calculate_Residual_Energy(i)
RETURN ω_1×p_1 + ω_2×p_2 + ω_3×p_3 + ω_4×p_4
```

Figure 5. Calculating fitness for CH selection

The distance parameter is very important as for Bluetooth the link quality will fade with the increasing of distance [25]. For the communication method of WLAN, the long communication distance will also have a negative effect to the throughput as well as the packet success rate [27]. Here, $nD$ in Eq. (12) is the normalized distance between each CH and its RNs and between each CH and the base station. The parameter $K$ stands for the number of clusters and $S_i$ is the number of RNs in $i_{th}$ cluster.

$$nD = \frac{\sum_i^m \left[ \sum_j^{m_i} d(RN_j, CH_i) + d(CH_i, S) \right]}{\sum_k^n d(N_i, S)} \tag{12}$$

As the CHs have to maintain its RNs in the cluster and server as communication hub between its RNs and the base station, they will cost more energy than the regular nodes, so we must ensure the selected CHs have relatively high residual energy. In the application of m-learning, we consider two kinds of transmission: one is real-time transmission such as the live audio captured by the microphone or live video captured by the camera, the other kind is off-line files such as the multimedia which has already been recorded before class and some text files which may contain questions from the teacher, answers or feedbacks from the learners. We use the same transmission model as [11] that the transmit/receive power of both WLAN and Bluetooth are constant within the $R_W/R_B$. The formula in Eq. (13) is used to compute the normalized residual energy that less value of $nE$ is popular.

$$nE = 1 - \frac{\sum_i^m E_{ni}}{m} \tag{13}$$

The selected CHs' mobility should be relatively small as CHs are essential for their RNs to communication with the base station so we must choose the nodes with low velocity as CHs to keep the clusters stable. In Eq. (14) $nM$ is the normalized value of sum of CHs' mobility, and $V_i$ stands for the current velocity of each CH which could be easily [33][34]. The CH can also manage its RNs through examining their connection regularly. A RN is deleted from the CH's RN list once the CH can't communicate with it. A node is permitted to join the CH if it moves into the communication radius of the CH and the number of RN is less than the maximum cluster size.

$$nM = \frac{1.0}{\sum_{i}^{m} V_i + 1.0} \quad (14)$$

It has been demonstrated that single-node cluster have a negative impacts on the network performance [24]. Meanwhile, the only node in the single-node cluster has to communicate with the base station by WLAN whose transmit power is much larger than the Bluetooth, we must make sure that the number of single-node cluster is as small as possible, and try to select the CHs that cover more nodes. The $N_{single}$ in Eq. (15) is the number of single-node cluster.

$$nSN = \frac{N_{single}}{n} \quad (15)$$

As Eq. (16) shows, the sum of the four weights, i.e. $\alpha$, $\beta$, $\gamma$ and $\delta$, is 1. They reflect the importance of the four parameters and can be adjusted under different application background. In some scenes with high mobility, $\gamma$ could be relatively bigger to select the nodes with low velocity as CHs. $\alpha$ could set a little bigger to the applications that need high communication quality.

$$\alpha + \beta + \gamma + \delta = 1 \quad (16)$$

**4.4 Cluster Maintenance**

As the CHs have to communicate with the base station using WLAN and at the same time, using Bluetooth as intra-cluster communication method, some of the CHs may run out of their power before re-clustering. Furthermore, the CHs from different clusters may consume their energy unevenly due to the diverse cluster sizes. To address these problems, we propose two methods of cluster maintenance before the next operation of re-clustering to prolong the group networking lifetime and balance the load among nearby CHs.

**4.4.1 CH Shift**

The idea of dynamic cluster adjustment is to divide interval between re-clustering operations into some tiny period, and each CH periodically calculates how long it can still be a CH. If a CH could not endure to the end of this tiny period, it should give up its role as CH and joins a new cluster to prolong the network lifetime or choose one of its RNs who has more energy as CH and let the other nodes join the new cluster.

This threshold is dynamically computed as the transmission speed is time varying, so we have to periodically calculate the lifetime of the CHs to see if their residual energy is sufficient for the next tiny period. In order to evaluate how much energy this period will consume, the energy consumption history is used as Eq. (17)

$$E_{current} = \alpha \times E_{last} + (1-\alpha) \times E_{average} \quad (17)$$

$E_{last}$ is the energy consumption of the last tiny period, and $E_{average}$ is the average energy consumption from the beginning of this cluster to the period before the last. $\alpha$ is a parameter between 0 and 1 used to adjust the weight. We assume this is the $k_{th}$ period, so the $E_{average}$ will be computed as Eq. (18) and the value of $E_{current}$ will be passed to $E_{last}$.

$$E_{average} = \frac{(k-2) \times E_{average} + E_{last}}{k-1} \quad (18)$$

If the residual energy of a CH is below the value of $E_{current}$, it could no longer be a CH. The CH shift of cluster maintenance will be triggered to prolong the lifetime of the CH as well as the lifetime of the network.

**4.4.2 RN Adjustment**

The CH will consume more energy if its cluster size is large. In case to adjust the energy consumption rate of nearby CHs to achieve the local load balance, we should shift some RNs to other clusters.

To figure out the average load of nearby CHs, a CH will send out the HELLO message periodically containing its personal ID, residual energy, $E_{current}$ as well as its location to the CHs within its communication range of WLAN. When a CH receives a HELLO message from another CH, it will process the message and add the worked information into the neighbor CH table including distance, ID and remaining time of CH.

The distance in the table is used to select the adjacent CHs, and remaining time of CH represents the load of a CH. We don't use the residual energy to estimate the load of a CH due to that the CHs of different cluster sizes will consume their residual energy unevenly so it may not be a real reflection of the CH's current load. Here we use the average CHs' remaining time as load description and low remaining time represents it's load is relatively high. The average remaining time is computed using Eq. (19)

$$T_{average} = \frac{1}{L}\sum_{i=1}^{L} RT_i \qquad i \in \left\{ 1 \leq i \leq L \,|\, Distance_i < R_N \right\} \qquad (19)$$

where $L$ is the length of the table and $RT$ represents remaining time as CH. We only consider the adjacent CHs whose distance between the current CH is less than $R_N$, e.g. 15m if taking the radius of Bluetooth into count. If the load of a CH is higher than the average which means that it has less remaining time as CH, it could release some of its RNs by let the RNs try to join another adjacent cluster to achieve the local load balance.

## 5. Performance Evaluation

In this section, we will evaluate the performance of our BeeCup protocol. We compare it against three well-known clustering protocols: PSO-GA [29], LEACH [10] as well as SEP [28] under the communication scenes of one-to-many and many-to-one as for the application of m-learning. We add the mobility management to all the protocols as our protocol. It is assumed that there are $N$ nodes randomly dispersed into two fields one of which is a rectangular area of $80\times80m^2$, and the other as shown in Fig. 6 is a region imitating the scene as in a classroom where there are more students in the front.

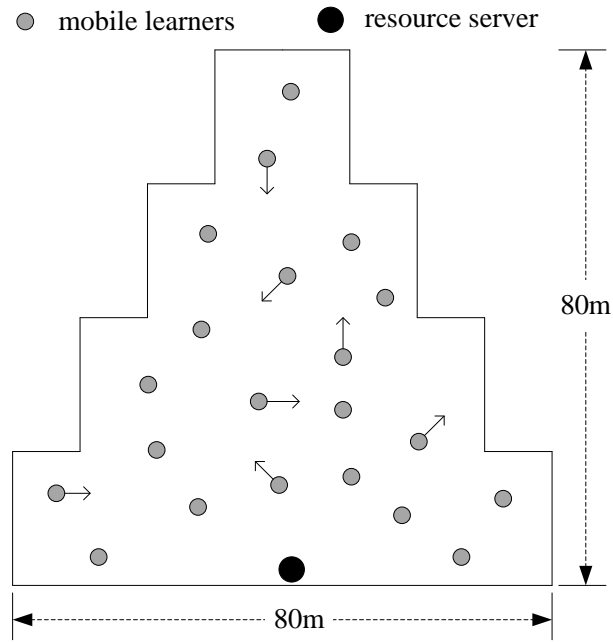

Figure 6. Simulated region imitating a classroom

The base station locates at the coordinate of (40, 0) in both regions. The number of nodes $N$ varies from 30 to 270 to test the performance of our protocol under different density. As the area of the second region is two-thirds of the first one, the value of $N$ for the second region is from 20 to 180.

We assume every node has two communication modules: WiFi and Bluetooth, whose communication radiuses are 100m and 10m respectively. We could use Bluetooth as intra-cluster communication method whose maximum cluster size is 7 and WLAN for communication between the CHs and base station. The communication parameters used in this simulation are the same as [11] and summarized in Table 3.

Table 3. Communication parameters

| Parameters | Values | Comments |
|---|---|---|
| $P^{W_A}$ | 1100mW | Active Tx/Rx power of WLAN |
| $P^{W_I}$ | 880mW | Idle listening power of WLAN |
| $P^{B_A}$ | 220mW | Active Tx/Rx power of Bluetooth |
| $P^{B_I}$ | 120mW | Idle listening power of Bluetooth |
| $R^W$ | 54Mbps | Maximum bit rate of WLAN |
| $R^B$ | 2Mbps | Maximum bit rate of Bluetooth |

There are two kinds of content to be communicated. The first is files which can be transmitted with the maximum bandwidth and the nodes turn to idle state after completion of transmission such as the recorded videos, images and so on. The second kind is real-time transmission such as the live audio captured by the microphone and live video captured by the camera, and nodes can only be turned to idle after the end of transmission of the source. In this simulation scene, we assume the total size of files transmitted is between 30MB and 60MB every 600s, and the time of real-time transmission is between 50s and 100s.

Unless otherwise specified, the simulation time is 7200s for each experiment and the re-clustering interval is set to 600s. The tiny period which is used for cluster maintenance is set as 60s. For the simulation of mobility, we use the Random Waypoint Movement with Pause model depicted in [35]. Each node randomly chooses a destination and a velocity between 0.5 and 1.0 m/s, and then moves to the destination. The node stays a randomly interval between 30 and 600 seconds after arriving at the destination and repeats the same procedure until the end of simulation.

We divide the evaluation of our BeeCup protocol into three parts: The first is to test the performance of CH number determination. In the second scene, the nodes are homogenous with the same initial energy of 10000J. The last is to test our clustering protocol under heterogeneous situation where the nodes are equipped with uneven initial energy.

**5.1 Determining Number of CHs**

First we evaluate the performance of our protocol in determining the optimal number of clusters under different density of nodes for both regions. The detailed results can be seen in Table 4 and Table 5, respectively. As shown in the tables, there are more single-node clusters when the density of nodes is low. This is because the limited number of nodes leads to more nodes which have no neighbors and thus have to communicate with the server directly. When there are more nodes in the area, the average cluster size grows towards to the maximum cluster size i.e. 7. As the second region is not as regular as the first one, some nodes may have less neighbors leading to more single-node clusters. We can see from Table 4 and Table 5 that the average cluster size in the first region is a little larger than the second one which confirms the validity of our clustering protocol's first part.

Table 4. Optimal number of clusters for region #1

| Number of nodes | Non-single cluster | Single-node cluster | Avg. cluster size |
|---|---|---|---|
| 30 | 8 | 8 | 2.75 |
| 60 | 15 | 9 | 3.4 |
| 90 | 20 | 7 | 4.15 |
| 120 | 25 | 3 | 4.68 |
| 150 | 28 | 4 | 5.21 |
| 180 | 31 | 3 | 5.71 |
| 210 | 34 | 1 | 6.15 |
| 240 | 37 | 2 | 6.43 |
| 270 | 40 | 1 | 6.73 |

Table 5. Optimal number of clusters for region #2

| Number of nodes | Non-single cluster | Single-node cluster | Avg. cluster size |
|---|---|---|---|
| 20 | 6 | 7 | 2.13 |
| 40 | 11 | 7 | 3.0 |
| 60 | 14 | 5 | 3.93 |
| 80 | 17 | 1 | 4.64 |
| 100 | 18 | 3 | 5.39 |
| 120 | 21 | 2 | 5.62 |
| 140 | 23 | 1 | 6.04 |
| 160 | 25 | 3 | 6.28 |
| 180 | 27 | 3 | 6.55 |

We can use the above results to guide the clustering process. When there are fewer clusters than the actual need, there will be many uncovered nodes. These nodes will resort to communicating with the server directly. However, when there are more nodes than the optimal number, the overhead of managing these clusters will outweigh their benefit. Thus, it is necessary to identify the optimal number of clusters to achieve BeeCup's best performance based on the distribution of the nodes.

**5.2 Clustering for Homogenous Nodes**

We use a set of experiments to show the capability of our clustering protocol on energy efficiency for homogenous nodes, as compared to PSO-GA and LEACH. Each performance index is simulated in both of the two regions. The same settings and assumptions are used for the three clustering protocols to make a reliable comparison. The initial energy is set to 10000J for all $N$ nodes. Simulation results are obtained by plotting of the mean values of more than 20 runs and each experiment uses a different randomly generated topology. Considering the application of m-learning, we set that 30% of all nodes per round have the mobility.

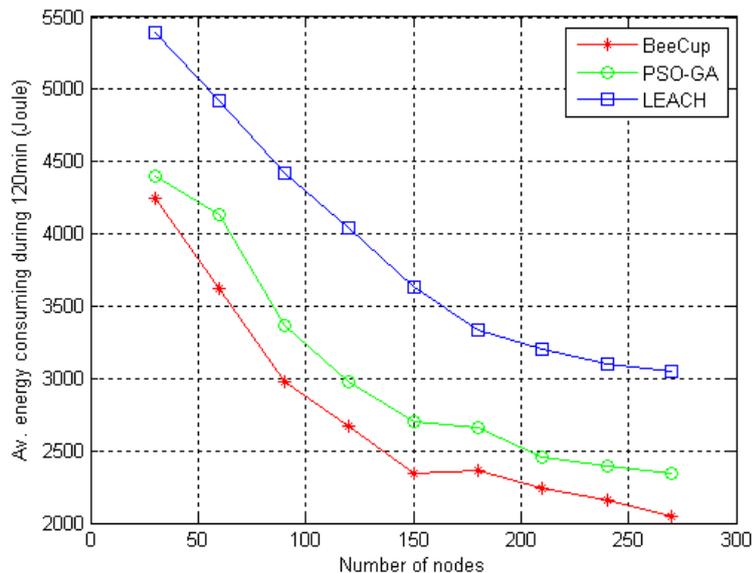

Figure 7. Average energy consumption during 120min for region #1

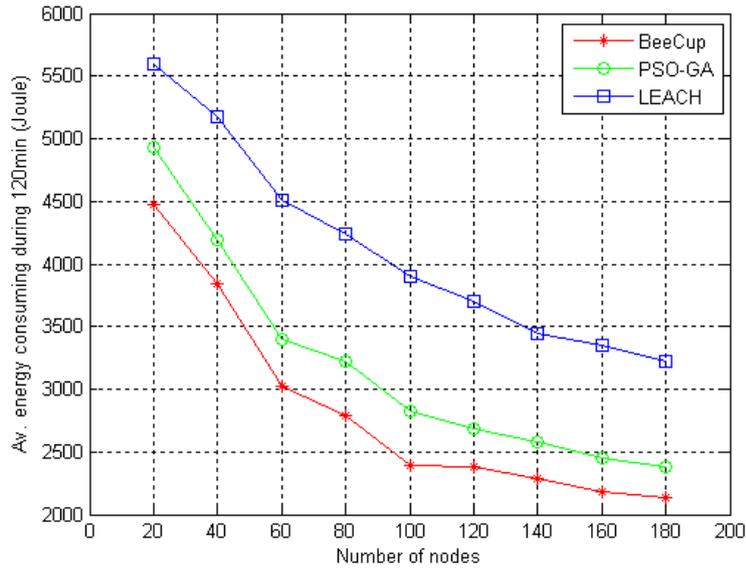

Figure 8. Average energy consumption during 120min for region #2

Fig. 7 and Fig. 8 show energy consumption for each node during 120 minutes of communication using PSO-GA, LEACH and our algorithm for two kinds of regions respectively. The number of nodes for the first region varies from 30 to 270, and for the second region it is from 20 to 180 in order to keep the same density.

We can see from the result that BeeCup clustering clearly reduces energy consumption more than LEACH and PSO-GA do for all situations. This is because LEACH randomly selects CHs and many nodes uncovered by these CHs have to communication with the server directly. For PSO-GA, the problem of single-node cluster is not well addressed. This is avoided by BeeCup as final CHs are selected such that they are well distributed and stable which ensure less existence of single-node cluster during a round. With the increase of nodes density, the average energy consumptions of LEACH and BeeCup descrease. This is because every CH will have more RNs around which don't need to communicate with the server.

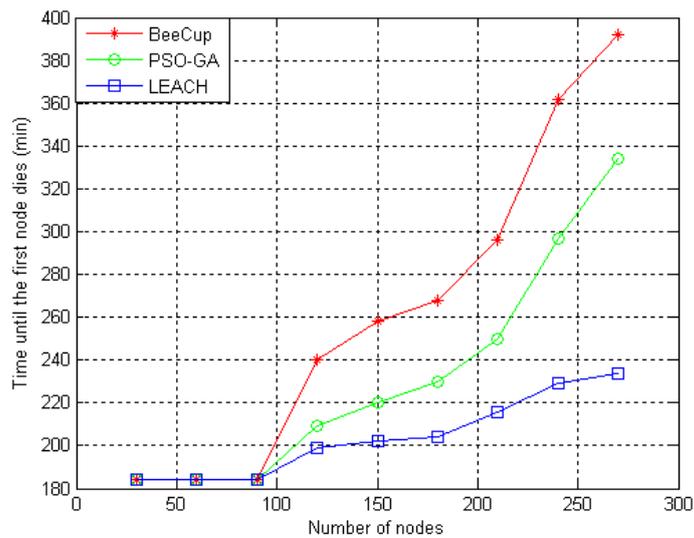

Figure 9. Time until the first node dies for region #1

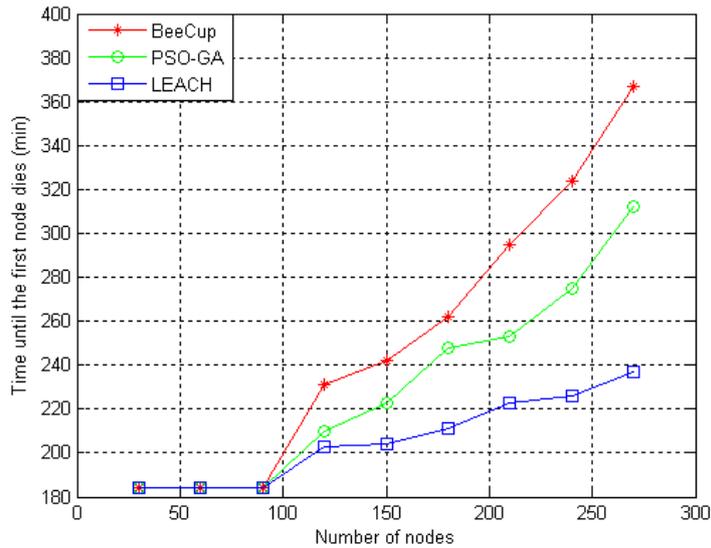

Figure 10. Time until the first node dies for region #2

A node is considered "dead" if it has lost 99.9% of its initial energy. The network lifetime is defined as the time until the first node dies. It is seen from Figs. 9 and 10 that when the number of nodes is small such as less than 90, the result is the same for the three methods. This is because as the distribution of nodes is sparse, there are always some single-cluster nodes which are out of other nodes' communication radiuses, and therefore they have to communicate with the server directly.

It is important to note that the network lifetime prolonged by BeeCup is especially significant with the number of nodes grows. This is because BeeCup can select the right nodes as CHs to reduce the number of single-node clusters and balance the energy consumption among the nodes. The cluster maintenance also contributes to the network lifetime when some nodes are not suitable to act as CHs due to their residual energy before re-clustering, or when the loads of the adjacent CHs are uneven. The situation of clusters can be dynamic adjusted with the help of cluster maintenance so that the nodes are always in their best roles.

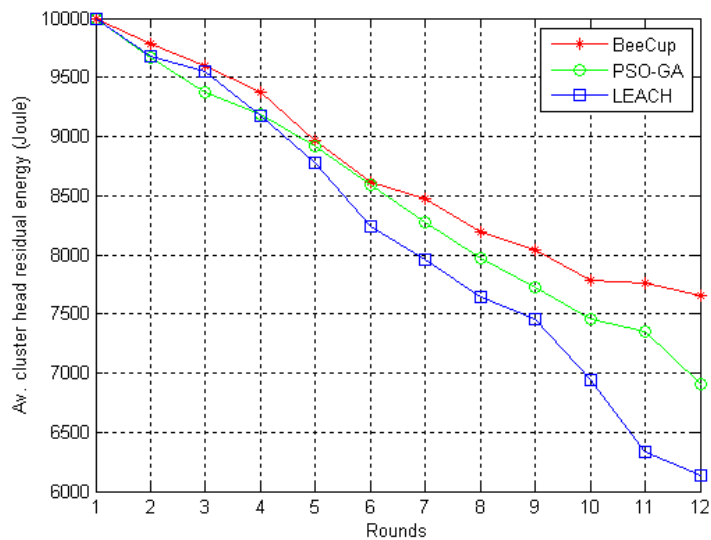

Figure 11. Average cluster head residual energy per round for region #1

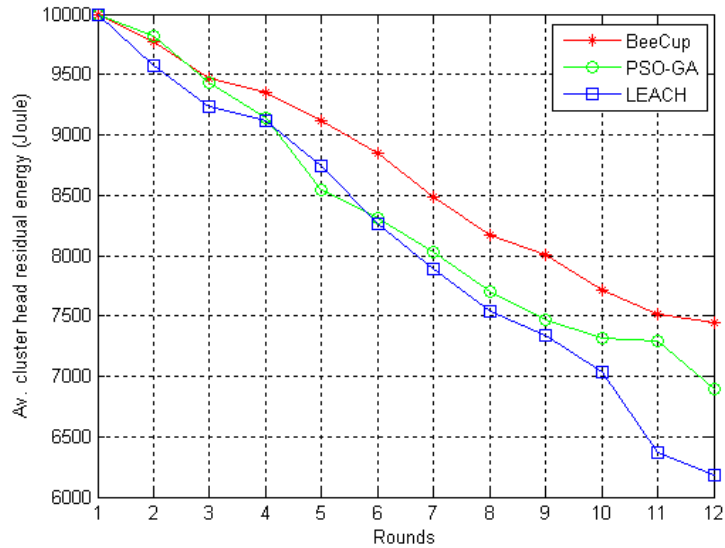

Figure 12. Average cluster head residual energy per round for region #2

As mentioned previously, BeeCup is load-balanced and hence can make energy consumption evenly among the nodes. This is proved by simulation results shown in Fig. 11 and Fig. 12. The simulation lasts for 120 minutes and the re-clustering interval is set to 10 minutes which is reasonable considering the energy consumption by CHs and the mobility. We sample the residual energy of CHs at the beginning of every round of both two regions. The results indicate that the selected cluster heads using BeeCup have more residual energy than the other two clustering protocols.

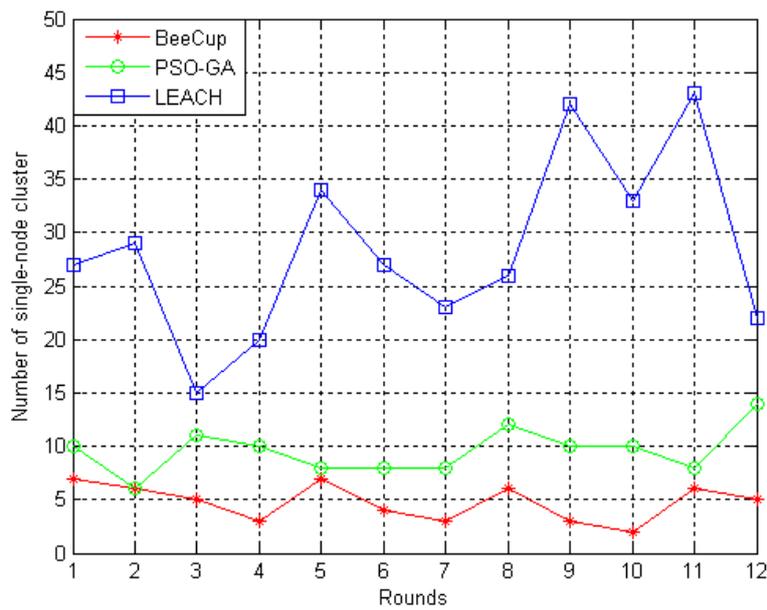

Figure 13. Number of single-node clusters per round for region #1

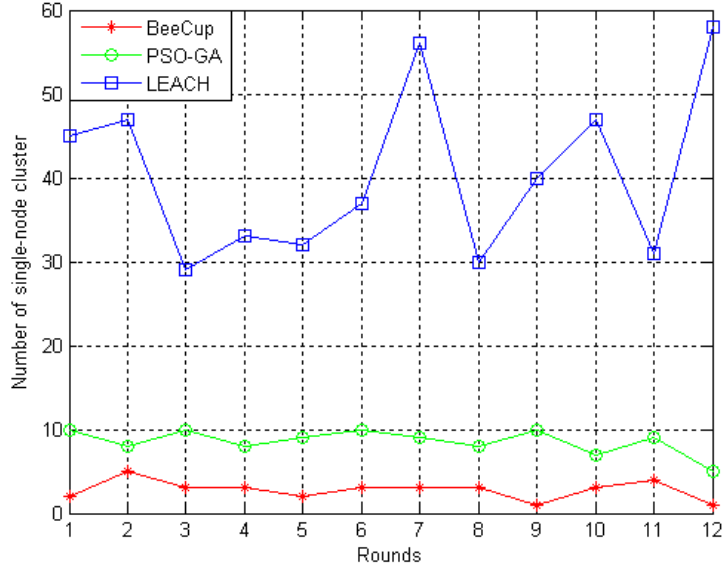

Figure 14. Number of single-node clusters per round for region #2

More single-node clusters will have negative impacts on the network performance [24], as these nodes have to communicate with the server directly, which will consume much energy on transmitting and receiving. Consequently, we adopt the number of single-node clusters as an important index in BeeCup and try to minimize this number. We use experiments to prove it in which we fix the number of nodes as 120 for the first region and 80 for the second region. The number of single-node clusters is sampled after each clustering process. The results are plotted in Fig. 13 and Fig. 14 respectively. It is shown that the number of single-node clusters is especially smaller using BeeCup than PSO-GA and LEACH. LEACH randomly selects the CHs whose distribution may be not very desirable and leave many nodes uncovered. This is improved in BeeCup and the distribution of CHs selected by our algorithm is very even.

**5.3 Clustering for Heterogeneous Nodes**

Next we compare our BeeCup protocol with a heterogeneous-aware clustering protocol, i.e. SEP [28]. SEP allows a percentage of all nodes to equip with extra initial energy than the normal nodes in the network. The parameter $m$ shows the fraction of advanced nodes and $\alpha$ is the additional energy factor between advanced and normal nodes. The thresholds for nodes to become the CHs are $T(s_{nrm})$ and $T(s_{adv})$ and can be computed as following in Eqs. (20)-(23):

$$p_{nrm} = \frac{p}{1+\alpha \times m} \qquad (20)$$

$$p_{adv} = \frac{p}{1+\alpha \times m} \times (1+\alpha) \qquad (21)$$

$$T(s_{nrm}) = \begin{cases} \dfrac{p_{nrm}}{1 - p_{nrm} \times \left(r \bmod \dfrac{1}{p_{nrm}}\right)} & \text{if} \quad s_{nrm} \in G' \\ 0 & \text{otherwise} \end{cases} \qquad (22)$$

$$T(s_{adv}) = \begin{cases} \dfrac{p_{adv}}{1 - p_{adv} \times \left(r \bmod \dfrac{1}{p_{adv}}\right)} & if \quad s_{adv} \in G'' \\ 0 & otherwise \end{cases} \quad (23)$$

where $r$ is the number of current rounds, $G'$ is the set of normal nodes that haven't been chosen as CHs during the last past $1/p_{nrm}$, and $G''$ is the set of advanced nodes that haven't become CHs within the last $1/p_{adv}$ rounds. According to the threshold for different kinds of nodes, the advanced nodes which have more residual energy will have higher probability to become CH that could efficiently prolong the network lifetime.

We compare the network lifetime under two groups of initial situations: first, 20% percentage of all nodes will have two times additional energy than the others; second, 20% percentage of all nodes have three times additional energy than the others. The results are shown in Figs. 15 and 16 for both the two regions respectively.

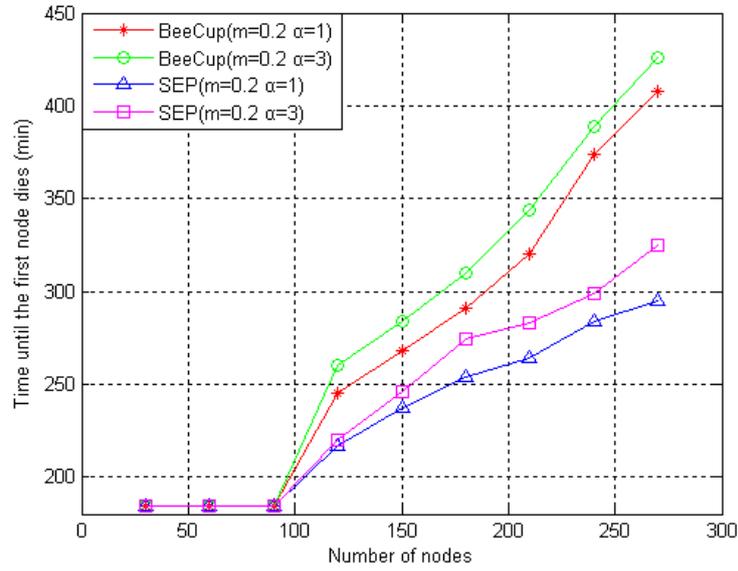

Figure 15. Network lifetime for region #1

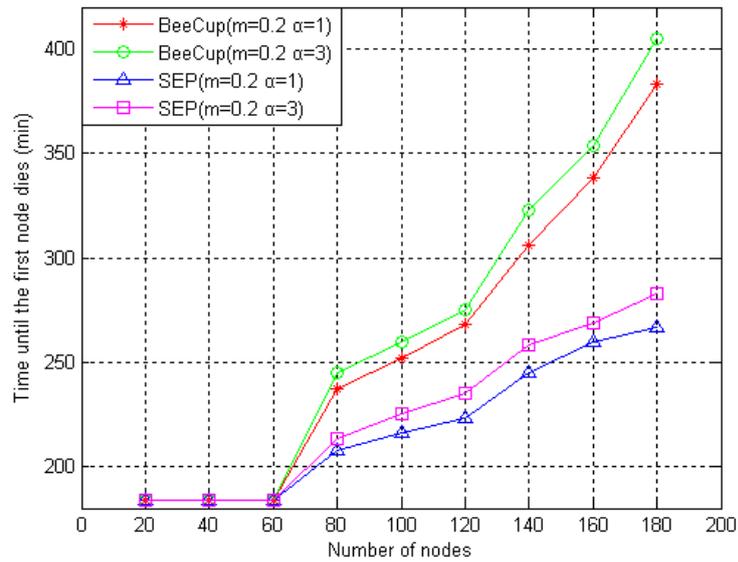

Figure 16. Network lifetime for region #2

From Fig. 15 and 16, we can see that the BeeCup can prolong the network lifetime as compared to SEP when the nodes have different initial energy. Moreover, the lifetime of network is longer when $\alpha$ is bigger. Even though SEP assigns different probabilities for the advanced and normal nodes, it still has high randomicity when selects the CHs and leaves the single-node cluster. Furthermore, by comparing this group of results with the homogenous situation, we can find out that the BeeCup protocol can prolong the network lifetime effectively when some nodes have addition energy. This is because BeeCup takes account of the factor of residual energy and selects the nodes with more energy under the same circumstances.

## 6. Conclusion

In this paper, we have presented BeeCup, a biologically-inspired energy-efficient and load-balanced clustering protocol, which can be used in the many-to-one and one-to-many communication scenes such as mobile learning. BeeCup first uses the bio-inspired algorithm ABC to adaptively determine the near-optimal number of clusters, and then selects the CHs taking into account the residual energy, mobility, number of single-node clusters as well as distance factors. The proposed clustering maintenance methods will adjust the clusters dynamically to keep the network structure stable. The performance of BeeCup has been evaluated under different network sizes and scenes. Our protocol was compared to three existing protocols: PSO-GA, LEACH, and SEP. The results show that BeeCup performs better in prolonging the network lifetime and hence can improve the quality of user experience in mobile learning. Furthermore, it is applicable to both homogenous and heterogeneous networks.

**Acknowledgements**

This work was partially supported by the Natural Science Foundation of China under Grants No. 60903153 and No. 61203165, the Fundamental Research Funds for Central Universities (DUT12JR10), and Liaoning Provincial Natural Science Foundation of China under Grant No.201202032.